# Galton's Family Heights Data Revisited

Hao HAN[a], Yeming MA[b], and Wei ZHU[a]

[a] *Department of Applied Math and Statistics, Stony Brook University, Stony Brook, NY*

[b] *Institutional Client Group, Citicorp, New York, NY*

Corresponding Email: hanhao224@gmail.com

**ABSTRACT**: Galton's family heights data has been a preeminent historical dataset in regression analysis, on which the original model and basic results have survived the close scrutiny of statisticians for 125 years. However by revisiting Galton's family data, we challenge whether Galton's classic model and his regression towards mean interpretation are proper. Using Galton's data as a benchmark for different regression methods, such as least squares, orthogonal regression, geometric mean regression, and least sine squares regression - a newly developed nonparametric robust regression approach, we elucidate that his regression model has fundamental drawbacks not only in variable and model selection by "transmuting" women into men thus the simple linear model, but also a strong bias in least squares regression leading to otherwise alternative conclusions on the true relationships between the heights of the child and his or her parents.

**KEYWORDS**: Galton's family (heights) data; errors-in-variables; least squares; orthogonal regression; geometric mean regression; least sine squares regression; regression efficiency.

## 1. INTRODUCTION TO GALTON'S 'REGRESSION'

The statistical terminology of 'regression' was coined by Sir Francis Galton beyond dispute, while the family heights data was formally introduced in his study on *Regression towards Mediocrity in Hereditary Stature* (Galton 1886, 1889). Whereas the original meaning of Galton's 'regression' has no significance in most of its applications (Mills 1924), it has been shown in the history of statistics that the regression method he proposed is just as important as the genetics laws he was trying to explain (Stigler 1986).

Galton's law of 'regression towards the mean' illustrated in Figure 1, that has been considered as one of the most appealing achievements in statistics, was originally described as follows.

When mid-parents are taller than average, their children tend to be shorter than they; when mid-parents are shorter than average, their children tend to be taller than they; the deviates of the children are to those of their mid-parents as 2 to 3 (Galton 1886).

Without knowledge of how Galton approached his "smoothed" sloping ratio, by adopting the well-known least squares (LS) method formulated in 1794, one observes the regression slope of child on mid-parent is indeed about 2 to 3. However, Galton's observation of this 'regression' phenomenon is questionable. Upon performing the inverse LS regression of mid-parent on child, one may draw an exactly opposite conclusion that the human stature is becoming more dispersed, when the slope of mid-parent on child is about 1/3 but not simply the reciprocal of 2/3.

The underlying simple yet subtle concept of regression towards the mean has repeatedly been the root of major errors in real-life applications (Stigler 1997). For instance, the most striking blunder by misapplying such concept happened on the book named *The triumph of mediocrity in business* (Secrist 1933), which was subsequently flogged that 'The seeming convergence is a statistical fallacy, resulting from the method of grouping (Hotelling 1933).' In general occasions, the regression fallacy – or, perhaps better, the regression trap could be easily committed in thinking that the phenomenon of regression towards the mean is due to certain nonrandom events, other than the nature of the football-shaped cloud of data points (Freedman, Pisani, Purves and Adhikari 1991).

## 2. BACKGROUND QUESTION AND METHODS

### 2.1 Raw Data

Fortunately, despite the elapsing of more than a century, the researchers were still able to retrieve Galton's family heights data from his firsthand notebook reserved at University College London (Hanley 2004a). It consists of the records from 205 families with 962 adult children in total, among which 486 are sons and 476 are daughters. However, after excluding the non-numerical entries (tall, medium, short, etc.), the preprocessed data in our article eventually consists of the records of 481 sons and 453 daughters as well as their parents. Given such elliptical scattered Galton's data, the evident divergence of different regression fits demonstrates the challenge to detect the true linear relationship behind the data (Figure 2).

## 2.2 Regression Models

Beyond all doubt, researchers are more concerning on the regression relationship of child on his/her parents instead of the inverse, which is of less scientific interest (Hanley 2005). Moreover, linear functions are assumed to be adequate to describe these models, in the sense that the nonlinear regression fits show no significant improvement over the linear one (Hanley 2004b), while a linear relationship is more straightforward to be interpreted.

More importantly, in reality, it is natural to raise a curiosity questioning whether the stature of the offspring inherits more from the father or the mother. To address this issue, we propose the pair of gender-specific multiple linear regression models (1) & (2) as we are interested in discriminating the model for sons from that for daughters.

$$Y_1 = \alpha_1 + \beta_{11}X_{11} + \beta_{12}X_{12} + \varepsilon_1 \quad (1)$$

$$Y_2 = \alpha_2 + \beta_{21}X_{21} + \beta_{22}X_{22} + \varepsilon_2 \quad (2)$$

The random variables of paternal height $X_{11}$ and the maternal height $X_{12}$ are bundled with the son's height $Y_1$ in model (1), while the daughter's height $Y_2$ together with the heights of her father $X_{21}$ and her mother $X_{22}$ are involved in model (2), where $\varepsilon_1$ and $\varepsilon_2$ are the corresponding error terms. The summary statistics for all the regression variables are tabulated in Table 1.

## 2.3 Regression Methods

The LS method has been almost universally adopted in the linear or nonlinear model estimation, often for predictive purposes. Unfortunately, it is unsuitable for the regression analysis of Galton's data in the following three folds. First, instead of constructing a regression model for the purpose of predicting the height of child given the parents' heights, we are primarily interested in accurately modeling the true functional relationship between the variables of interest. Second, even with predictive purposes, there is still a lack of prior knowledge to conventionally differentiate the response and explanatory variables in Galton's data (Los 1999), but the LS concerns on the prediction accuracy of the response variable exclusively. Last but not the least, due to the potential of operational errors or measuring instrument failures, the regression methods applicable to the measurement error model, also called errors-in-variables (EIV) model, are more desirable rather than the LS method, which gives inconsistent estimates in the presence of measurement errors (Cochran 1968, Casella and Berger 2002). Therefore, some alternative approaches such as orthogonal regression (OR) and geometric mean regression (GMR) (Draper and Yang 1997), which account for the errors in both the dependent and independent variables, will be considered here.

Furthermore, we will examine the performance of a novel robust regression method - least sine squares (LSS), proposed in our previous work (Han 2011, Han et al. 2012), to the analysis of Galton's data. The LSS, after its name, makes use of an angular measure of $sin(\varphi_i)$. In the two dimensional case, $\varphi_i$ represents the angle formed by the fitted regression line and the line connecting each data point with the center of the dataset. Different from minimizing the sum of squared orthogonal distances in OR, the objective of LSS is to minimize the sum of squared *sine* distances $\sum_i sin^2\varphi_i$, and the slope estimator for the simple LSS regression is $\hat{\beta} = \frac{\tilde{S}_{YY}-\tilde{S}_{XX}+\sqrt{(\tilde{S}_{YY}-\tilde{S}_{XX})^2+4\tilde{S}_{XY}^2}}{2\tilde{S}_{XY}}$ with $\tilde{S}_{YY} = \sum_i \frac{(Y_i-\bar{Y})^2}{R_i^2}, \tilde{S}_{XX} = \sum_i \frac{(X_i-\bar{X})^2}{R_i^2}, \tilde{S}_{XY} = \sum_i \frac{(X_i-\bar{X})(Y_i-\bar{Y})}{R_i^2}$, where $R_i$ is the distance from each point $(X_i, Y_i)$ to the centroid $(\bar{X}, \bar{Y})$. The conceptual representations of different methods of interest are illustrated through Figure 3.

## 3. ANALYSES

### 3.1 Different Regression Fits

The estimated regression coefficients from different approaches (Table 2) clearly demonstrate that the LS slope estimates are always much smaller than that from the other regressions configured for EIV models, due to the nature that the LS will underestimate the regression slopes when the predictors are contaminated with measurement errors (Figure 4). Of note, referring to the estimated regression slopes for the model on daughters' heights, the LS slopes relative to the others are in a reverse pattern. Explicitly, since the daughter-father's partial correlation as of 0.444 is larger than the daughter-mother's as of 0.329, and relative to the father's standard error as of 2.65 the mother's as of 2.26 is not small enough, the regression coefficient of the father term is bound to be larger than that of the mother term.

### 3.2 Goodness-of-fit for Each Method

As we have seen, different methods lead to distinct estimates of regression coefficients for the given dataset, but which estimation is the most suitable one? Recalling the diagnostics of the least squares regression models, the well-known $\chi^2$ goodness-of-fit test or the coefficient of determination – $R^2$ is often used as an indicator to gauge how much the variation in the original data has been explained by the fitted regression model. Unfortunately, both indicators are confined to assess the adequacy of least squares estimation only. To compare the LS fit with the other non-LS fits for gender-specific models, we define the regression efficiency with respect to each involved variable as follows.

$$e_Y = \frac{min \sum_{i=1}^n (Y_i - \tilde{Y}_i)^2}{\sum_{i=1}^n (Y_i - \bar{Y}_i)^2}, \quad e_{X_1} = \frac{min \sum_{i=1}^n (X_{1i} - \tilde{X}_{1i})^2}{\sum_{i=1}^n (X_{1i} - \bar{X}_{1i})^2}, \text{ and } e_{X_2} = \frac{min \sum_{i=1}^n (X_{2i} - \tilde{X}_{2i})^2}{\sum_{i=1}^n (X_{2i} - \bar{X}_{2i})^2}$$

The regression efficiency with respect to the specific variable is calculated as the ratio of the minimized to the observed sum of squared residuals along the associated coordinate direction, and it gauges how optimized the fitted model is in minimizing the prediction error of that particular variable. Intuitively, since all the variables are equally important, the higher the sum of regression efficiencies (SRE), the better the fitted model is. Compared with the other estimators designated for EIV model estimation, the LS has much lower SREs (Table 2), which to some extent reflects that not only the response variables in our regression models are contaminated with measurement errors.

Since it has been proven that the GMR always has the highest SRE with equal regression efficiencies w.r.t. each variable when the data follows a bivariate normal distribution, it is natural to propose the conjecture that the GMR should also attain the highest SRE with equal regression efficiencies even for the multivariate case. Based on current data analysis, this conjecture holds for the model on daughter's height, but for the model on son's height the OR not the GMR has the highest SRE, which implies that there must be some underlying assumptions if the conjecture were valid. Moreover, as the multivariate GMR estimation is solved through iterated fractional programming and there is no closed-form estimator, it is therefore necessary to numerically examine the optimality of each regression method in multivariate case by means of simulation studies.

### 3.3 Simulation Studies of Regression with Two Predictors

The analyses on real data set may lead to disputes, because it is impossible to 'prove' which method gives the best estimate without knowing the truth. A popular alternative way is to carry out Monte Carlo simulations, as one always knows the true parameters behind the constructed data set.

Since there are no gross outliers diagnosed through either the LS residuals or some robust technique for Galton's data, our simulation experiment merely focused on the study of regressions when variables were contaminated with measurement errors. To simply serve our purpose, the experiment was designed for the case of regressions with two predictors only that is comparative with our gender-specific models.

For each constructed data set $(X_1, X_2, Y)$, the true underlying random variables were firstly generated as $\xi_i \sim N(0, 100)$ for $i = 1, 2$, and then w.l.o.g. $\eta$ is given by evaluating a trivial linear relationship as $\eta = \beta_0 + \beta_1 \xi_1 + \beta_2 \xi_2$ when $\beta_0 = \beta_1 = \beta_2 = 1$. The independent normal measurement errors were then added to each of the true values to obtain a set of observed $X_1 = \xi_1 + \delta_1$, $X_2 = \xi_2 + \delta_2$, and $Y = \eta + \varepsilon$ values, and we have basically resorted to four types of configurations of the error (measurement error) variances:

1. small error, equal error variances in $(X_1, X_2, Y)$;
2. small error, unequal error variances in $(X_1, X_2, Y)$;
3. large error, equal error variances in $(X_1, X_2, Y)$;
4. large error, unequal error variances in $(X_1, X_2, Y)$.

Here "small" means that the error variance is smaller than 10% of the variance of the underlying true but unobserved $\xi$'s or $\eta$; "large" means that the error variance is larger than 10%, but less than 50%, of the variance of the underlying true but unobserved $\xi$'s or $\eta$ (Draper, et al. 1997).

We considered 1000 samples for each choice of sample size $n = 500$ (large sample) and $n = 50$ (small sample); and for all the four error variances situations. In order to compare the performance of the four different regression methods, several estimators were investigated on the same simulated data set respectively. The purpose of our simulation is to measure to what extent the estimates differ from the true values of regression coefficients $\beta_0 = \beta_1 = \beta_2 = 1$. Some summary values over 1000 runs are computed, such as the mean estimated value, the mean squared error, the variance, and the p-value of hypothesis test on the true value of each regression coefficient. To compare the optimality of each method, we used the total mean squared error (TMSE) of all three coefficient estimates as well as that of only the two slope estimates that we are most interested in as the criterion of goodness-of-estimation.

Table 4(a) gives the simulation results under the configuration of a large sample size of $n = 500$ with small errors. We see that in the equal variances situation the mean estimated values produced by both OR and GMR are not significantly different from the true coefficients at the significance level of 0.05, while the LS and LSS have significantly ($p < 10^{-6}$) biased slope estimates. It may be noted that the constant term produces the largest MSE and variance for all four methods. In terms of TMSE, the GMR provides the best estimation and the OR also behaves quite well, while the novel LSS performs better than the ordinary LS. Moreover, for the situations of unequal variances, the GMR still has the smallest TMSE, whereas the slope estimates from all methods are significantly biased. Similarly, for the results tabulated in Table 4(b) based on the setting of a large sample size with large errors, we arrives the same conclusions as before. To conclude, in the terms of TMSE w.r.t. no matter all regression coefficients or only the two slopes, the GMR approach is generally the optimal one and the OR is the near optimal one but with a bigger variance for the large sample EIV problems.

On the other hand, Tables 4(c) and 4(d) demonstrate the results of the simulation experiment conducted under a small sample size of $n = 50$. It is necessary to mention that for almost all the situations constructed here, albeit the LS estimations compared to others have the smallest TMSE w.r.t. all coefficients, it is the GMR but not the LS which has the smallest TMSE w.r.t. the slopes. This discrepancy between the two different forms of TMSE was introduced by the relatively small variance of the LS intercept estimate, and the relatively large bias of the LS slope estimates in small sample situations.

Back to our aim on the verification of previous conclusions on Galton's data, we ought to concentrate more on the simulation results of large sample scenarios, and the accuracy of slope estimates i.e. the TMSE w.r.t. slope estimates is definitely of more interest. Therefore, the rule of thumb from these simulations leads us to in favor of the GMR approach even if dealing with a small sample problem, while compared to the ordinary LS, the OR as well as the LSS is the near optimal ones which also provide decent estimations of the regression model. It is as expected consistent with our previous insights of the optimality of each regression method based on the novel concept of regression efficiency.

**3.4  Hypotheses Testing on Regression Slopes**

After evaluating the goodness-of-fit of each approach, we also care about whether the identified patterns of unequal contributions from parents are statistically significant or merely occur by chance. Due to the violation of the normality assumption for the gender-specific models (Table 1), our precedent insight of parametric hypotheses testing is questionable. Meanwhile, even if the underlying assumptions for parametric tests are fulfilled, it is still complicated for the inference based on the asymptotically estimated covariance matrix for regression coefficients of multivariate EIV models (Patefield 1981). Hence, we will take advantage of the prevailing non-parametric technique - the bootstrap (Efron 1979, 1982, Efron and Tibshirani 1993) to test the hypotheses.

Presumably, the parents are in equal roles to the stature of the offspring, our null hypotheses would be $H_{01}$: $\beta_{11} = \beta_{12}$ for the model on the sons' heights, and $H_{02}$: $\beta_{21} = \beta_{22}$ for the model on the daughters' heights respectively. Under each null hypothesis, since both terms from father and mother in the regression model will be ultimately merged into one single term, it is therefore feasible to set up a permutation test by randomly swapping the father's and the mother's heights within each family. Furthermore, a generalized bootstrap resampling strategy is also utilized by combining the bootstrap of the regression pairs ($Y_i$, $X_i$) with the random swap of paternal and maternal heights within each bootstrapped pair.

If the observed positive differences of slopes in gender-specific models are denoted as $\hat{\Delta}_1 = \hat{\beta}_{11} - \hat{\beta}_{12}$ and $\hat{\Delta}_2 = \hat{\beta}_{22} - \hat{\beta}_{21}$ respectively, the corresponding resampled differences of $\hat{\Delta}_1^* = \hat{\beta}_{11}^* - \hat{\beta}_{12}^*$ and $\hat{\Delta}_2^* = \hat{\beta}_{22}^* - \hat{\beta}_{21}^*$ will then be hypothetically generated through the resampling procedures. Consequently, the achieved significance level (ASL) of each hypothesis test is defined to be the probability of observing at least that large a difference when the null hypothesis is true, and the corresponding ASLs are formulated as $\text{ASL}_{Son} = P_{H_{01}}(\hat{\Delta}_1^* \geq \hat{\Delta}_1)$, and $\text{ASL}_{Daughter} = P_{H_{02}}(\hat{\Delta}_2^* \geq \hat{\Delta}_2)$.

It is not surprising that the ASLs from two different resampling procedures are quite similar (Table 3), but we are more confident in the results acquired after the family wise reshuffling through the generalized bootstrap. Excluding the hypothesis testing on the irrational LS estimates, the OR, the GMR, and especially the LSS estimations have justified the significance of the observed patterns of unequal slopes, which are coherent for all the three regression methods.

## 4. DISCUSSION

In retrospect, the divergence between our analysis and Galton's regression starts from the step of variable selection. Although the sons are not significantly different from the fathers, nor the daughters and the mothers, the male subjects with significantly larger sample variance ($p = .0018$) are in average significantly taller than the females ($p < .0001$). By taking account of this heterogeneity across genders, Galton proposed to transmute the height of each female to the male equivalent by multiplying by a factor of 1.08 (Galton 1886). As a result of this magic manipulation, the heterogeneity was diminished to an acceptable level, and it is then reasonable for him to treat the unisex child (son/daughter) as the response variable, and the mid-parent (average height of father and mother) as the single predictor.

Whereas the gender-specific models presented in this article had not been appreciated by Galton, the essentiality of considering gender differences seemed attractive to him.

I use the word parent to save any complication due to a fact apparently brought out by these inquiries, that the height of the children of both sexes, but especially that of the daughters, takes after the height of the father more than it does after that of the mother. My present data are insufficient to enable me to speak with any confidence on this point, much less to determine the ratio satisfactorily (Galton 1886).

In essence, Pearson substantiated Galton's insight by the pairwise simple regressions and correlations from Galton's data (Pearson 1896). However, by collecting a much larger data series of British familial heights, Pearson found all the pairwise correlations to be approximately .50, and the coefficients of the mother's height in the least squares estimated multiple linear equations were invariably higher than the father's coefficients (Pearson and Lee 1903). The contradicted conclusions were later attributed to the mismeasurement of statures in women for Galton's data (Pearson 1930). When we set aside Pearson's inspection of Galton's classic model, we clearly see that the family heights data was preferably generation and gender classified in Pearson's subsequent analyses.

Moreover, the discovered nonlinearity in both Galton and Pearson's regression of mid-parent on child again elucidated that the pooling of gender blocks was improper, which could be explained that, with child height as the covariate, the test for homogeneity of slopes with respect to parent (mother/father) was significant ($p < .01$) (Wachsmuth, Wilkinson and Dallal 2003).

However, Hanley argued that the nonlinear regressions, quadratic and cubic, did not significantly improve the regression fit of child on mid-parent over the linear one. When comparing the sharpness of different strategies for dealing with the fact that sons are generally taller than daughters and continuing to use Galton's definition of mid-parent, Hanley's sex-specific simple regression models, together with Galton's classic model after applying either the multiplicative strategy or the modern-day blackbox approach by adding 5.2 inches to each daughter's height, showed very close results with respect to correlations, regression slopes, and root mean squared errors. For the choice of regression method, Hanley specifically stressed that the least-squares regression was used in order to narrow the now-versus-then comparison (Hanley 2004b).

Instead of pursing Galton's inappropriate route of regression analysis, the aim of our revisit is to reveal the mechanism of how the offspring's stature inherits from each of the parents. We concluded that the stature of the son is mainly contributed from his father, while the stature of the daughter resembles her mother more closely rather than her father. Since our results are presented under the gender-specific multiple regression models, we ought to emphasize that the variable selection as well as the specification of a proper model are the critical steps to identify the alternative conclusions that the other models cannot attain. Furthermore, it also lies on the utilization of suitable regression methods for the model estimation.

Interestingly, according to the renowned half-half chance model in genetics that each parent contributes equally to the genetic makeup of their offspring, Clemons argued that it is also true as far as human stature is concerned. Inspired by Pearson and Lee's results, he drew a conclusion that the mother's measurements were more important than the reported father's measurements, as the mother's height was more accurately measured than the father's. Henceforth, he proposed a model that assigned the father term with a probability of incorrect measurement, and the equality of regression coefficients from parents could be finally realized by adjusting that probability for Pearson's data (Clemons 2000). However, from our point of view, Clemons's argument that was accomplished by artificially tuning the parameter of probability is indefensible, in the sense that technically the true probability of measurement errors is unknown, and psychologically the statisticians' prior expectations influence their posterior judgments.

# REFERENCES

**Casella, G., and Berger, R. L. (2002),** *Statistical Inference* **(2nd edition. ed.), Pacific Grove, CA: Duxbury.**

**Clemons, T. (2000), "A Look at the Inheritance of Height Using Regression toward the Mean,"** *Human Biology***, 72, 447-454.**

**Cochran, W. G. (1968), "Errors of Measurement in Statistics,"** *Technometrics***, 10, 637-&.**

**Draper, N. R., and Yang, Y. H. (1997), "Generalization of the Geometric Mean Functional Relationship,"** *Computational Statistics & Data Analysis***, 23, 355-372.**

**Efron, B. (1979), "Bootstrap Methods: Another Look at the Jackknife,"** *Annals of Statistics***, 7.**

**Efron, B. (1982)**, *The Jackknife, the Bootstrap, and Other Resampling Plans*, Philadelphia, Pa.: Society for Industrial and Applied Mathematics.

**Efron, B., and Tibshirani, R. (1993)**, *An Introduction to the Bootstrap*, New York: Chapman & Hall.

**Freedman, D., Pisani, R., Purves, R., and Adhikari, A. (1991)**, *Statistics* (2nd ed.), New York: W. W. Norton & Company, Inc.

**Galton, F. (1886), "Regression Towards Mediocrity in Hereditary Stature,"** *Journal of the Anthropological Institute of Great Britain and Ireland*, 15, 246-263.

**Galton, F. (1889)**, *Natural Inheritance*, London,: Macmillan.

**Han, H. (2011)**, *Least Sine Squares and Robust Compound Regression Analysis*, Stony Brook Theses & Dissertations, The Graduate School, Stony Brook University: Stony Brook, NY

**Han, H., Ma, Y., Jiao, X., Leng, L., Liang, Z., and Zhu, W. (2012), "Robust Compound Regression: A New Approach for Robust Estimation of Errors-in-Variables Models," In** *JSM Proceedings*, Nonparametric Statistics Section. Alexandria, VA: American Statistical Association.

**Hanley, J. A. (2004a), "Galton's Family Data on Human Stature,"** <http://www.medicine.mcgill.ca/epidemiology/hanley/galton/>.

**Hanley, J. A. (2004b), ""Transmuting" Women into Men: Galton's Family Data on Human Stature,"** *The American Statistician*, 58, 237-243.

**Hanley, J. A. (2005), "Reply to Comment of 'Transmuting' Women into Men: Galton's Family Data on Human Stature, by Wilkinson, Wachsmuth, and Dallal,"** *The American Statistician*, 59, 1.

**Hotelling, H. (1933), "Review of the Triumph of Mediocrity in Business by Horace Secrist,"** *Journal of the American Statistical Association*, 28.

**Los, C. A. (1999), "Galton's Error and the under-Representation of Systematic Risk,"** *Journal of Banking and Finance*, 23, 1793-1829.

**Mills, F. C. (1924)**, *Statistical Methods Applied to Economics and Business*, New York,: H. Holt.

**Nesselroade, J. R., Stigler, S. M., and Baltes, P. B. (1980), "Regression toward the Mean and the Study of Change,"** *Psychological Bulletin*, 88, 622-637.

**Patefield, W. M. (1981), "Multivariate Linear Relationships - Maximum-Likelihood Estimation and Regression Bounds,"** *Journal of the Royal Statistical Society Series B-Methodological*, 43, 342-352.

**Pearson, K. (1896),** "Mathematical Contributions to the Theory of Evolution. Iii. Regression, Heredity, and Panmixia," *Philosophical Transactions of the Royal Society of London*, 187, 253-318.

**Pearson, K. (1930),** *The Life, Letters and Labours of Francis Galton* (Vol. 3), Cambridge: Cambridge University Press.

**Pearson, K., and Lee, A. (1903),** "On the Laws of Inheritance in Man: I. Inheritance of Physical Characters," *Biometrika*, 2, 357-462.

**Secrist, H. (1933),** *The Triumph of Mediocrity in Business*, Evanston, Ill.: Bureau of Business Research, Northwestern University.

**Stigler, S. M. (1986),** *The History of Statistics: The Measurement of Uncertainty before 1900*, Cambridge, MA: Harvard University Press.

**Stigler, S. M. (1997),** "Regression Towards the Mean, Historically Considered," *Stat Methods Med Res*, 6, 103-114.

**Wachsmuth, A., Wilkinson, L., and Dallal, G. E. (2003),** "Galton's Bend: A Previously Undiscovered Nonlinearity in Galton's Family Stature Regression," *The American Statistician*, 57, 190-192.

# FIGURES AND TABLES

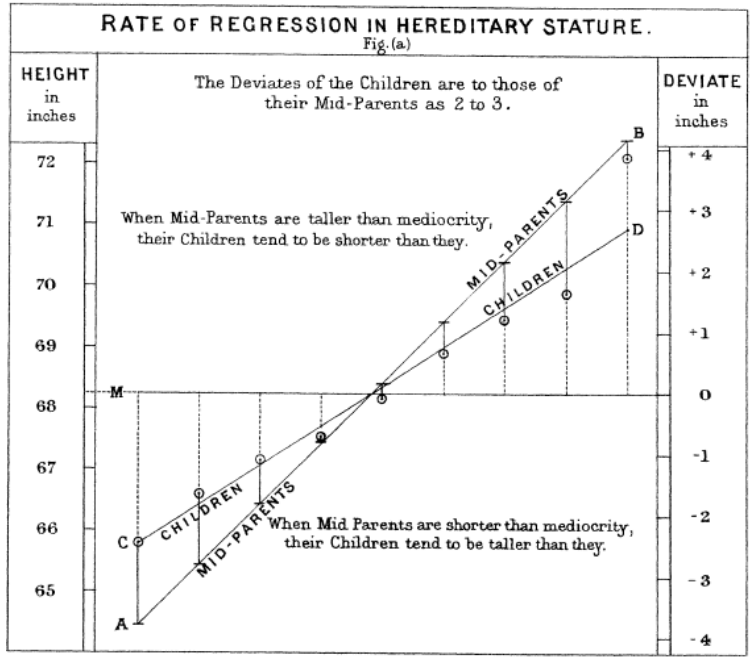

*Figure 1. Rate of regression in hereditary stature (Galton 1886 Plate IX, fig. a). The short horizontal lines refer to the height of mid-parents in inches from 64.5 to 72.5 with a step of 1. The small circles show the median height of the children of each of those mid-parents. The line AB passes through all mid-parental heights, and the line CD gives the "smoothed" results of the corresponding median heights of their children. The ratio of CM to AM is as 2 to 3. The point of convergence is at the level of mediocrity M = 68.25 inches. Note that all female heights have been converted to their male equivalents by multiplying each of them by 1.08.*

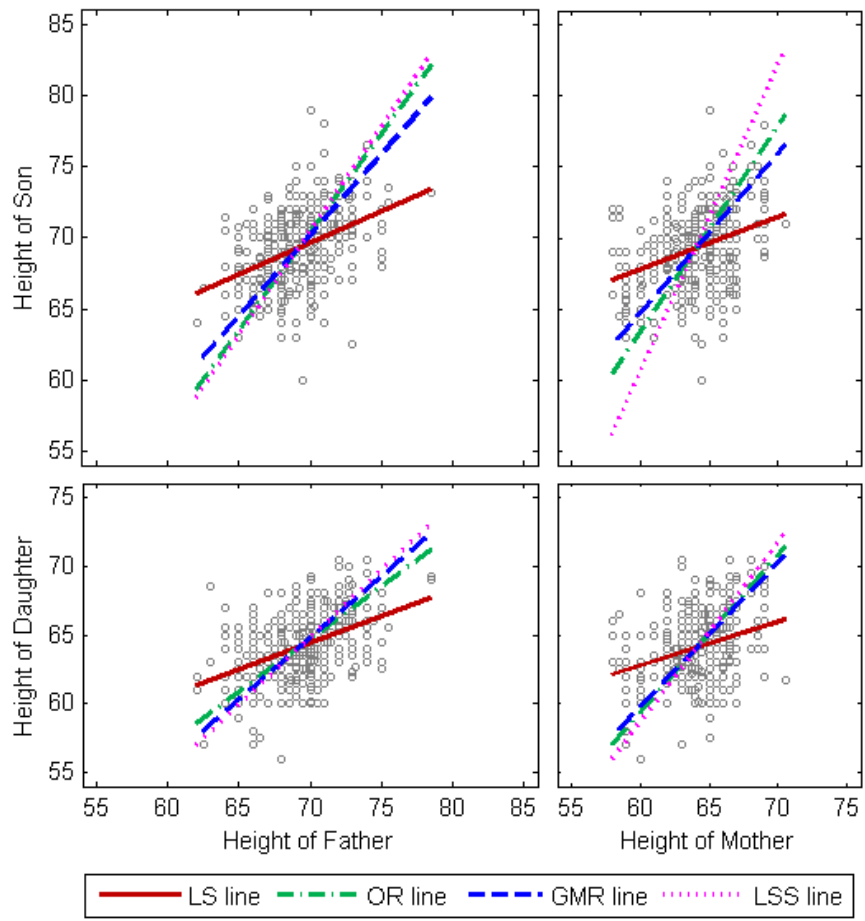

*Figure 2. Full gender cross-tabulation of Galton's family data with different regression fits. The* LS *lines (red solid) are obviously downward tilted away from the other regression lines.*

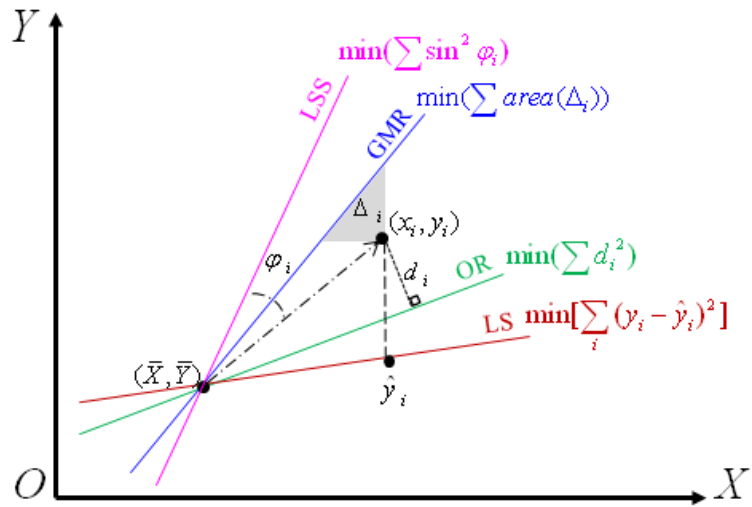

*Figure 3. Conceptual representations of different methods of interest. The* LS *regression minimizes the sum of squared vertical distances from the observation point to the regression line. The* OR *takes the middle ground by minimizing sum of squared orthogonal distances. The* GMR *minimizes sum of the triangular areas bounded by the regression line and the vertical and horizontal lines through each observation point. The novel* LSS *explores the best line by minimizing sum of squared sine of the angle formed by the fitted regression line and the line connecting each data point with the center of the dataset.*

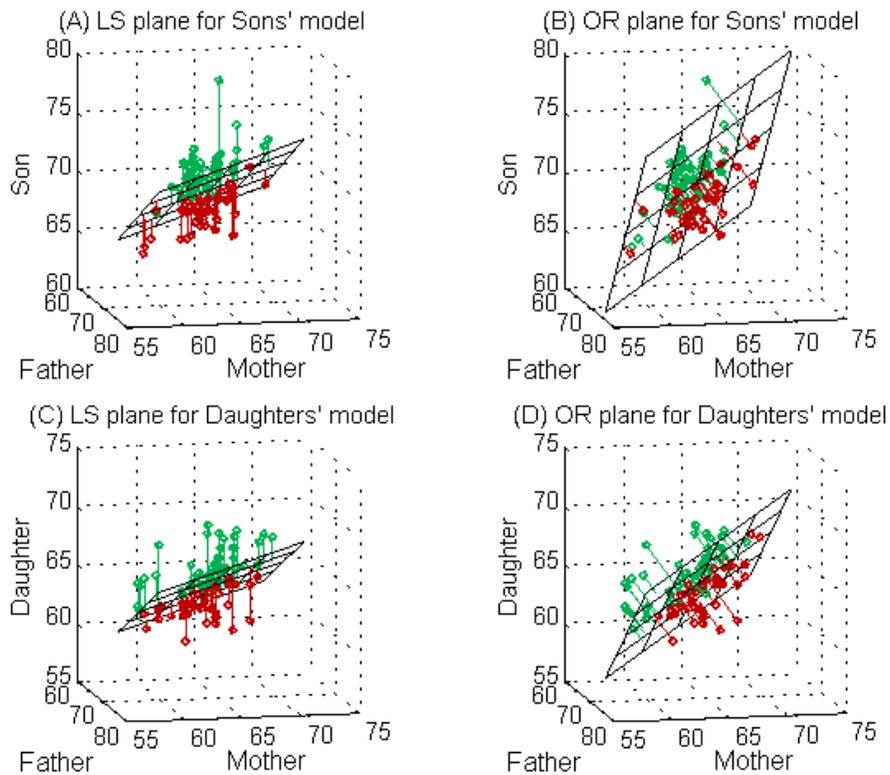

*Figure 4. The fitted planes from* LS (*Panels* A & C) *and* OR (*Panels* B & D) *approaches for the gender-specific models. Compared with* OR *fitted planes, the* LS *fitted planes do not explain the football-shaped cloud of data points very well, because the* LS *regression slopes are underestimated in* EIV *situations and thus the downward tilted planes. Of note, only 100 points were randomly chosen to be plotted in each panel for a clearer view.*

*Table 1. Summary statistics of gender-specific models for Galton's family heights data*

| Summary statistics | $Y_1 = \alpha_1 + \beta_{11}X_{11} + \beta_{12}X_{12} + \varepsilon_1$ | | | $Y_2 = \alpha_2 + \beta_{21}X_{21} + \beta_{22}X_{22} + \varepsilon_2$ | | |
|---|---|---|---|---|---|---|
| | $Y_1$ | $X_{11}$ | $X_{12}$ | $Y_2$ | $X_{21}$ | $X_{22}$ |
| Mean | 69.23 | 69.14 | 64.03 | 64.10 | 69.26 | 64.16 |
| Standard error | 2.62 | 2.31 | 2.32 | 2.36 | 2.65 | 2.26 |
| Skewness | -0.04 | 0.11 | -0.29 | 0.02 | 0.13 | -0.32 |
| Kurtosis | 3.28 | 3.51 | 3.02 | 3.20 | 3.50 | 3.26 |
| Shapiro-Wilk test | 0.0054 | 0.0001 | <0.0001 | 0.0069 | 0.0001 | <0.0001 |

Note: The statistics of mean and standard error are in inches.

*Table 2. Results from different regressions for gender-specific models*

| Methods | $Y_1 = \alpha_1 + \beta_{11}X_{11} + \beta_{12}X_{12} + \varepsilon_1$ | | | | $Y_2 = \alpha_2 + \beta_{21}X_{21} + \beta_{22}X_{22} + \varepsilon_2$ | | | |
|---|---|---|---|---|---|---|---|---|
| | $\alpha_1^*$ | $\beta_{11}^*$ | $\beta_{12}^*$ | SRE* | $\alpha_2^*$ | $\beta_{21}^*$ | $\beta_{22}^*$ | SRE* |
| LS | 19.3314 (.0533) | 0.4170 (.0006) | 0.3291 (.0006) | 1.2509 (.0005) | 18.1968 (.0547) | 0.3750 (.0005) | 0.3107 (.0006) | 1.3078 (.0007) |
| OR | -43.3346 (.1510) | 0.9769 (.0021) | 0.7032 (.0021) | 1.5125 (.0008) | -21.9101 (.1020) | 0.6147 (.0010) | 0.6771 (.0015) | 1.6001 (.0008) |
| GMR | -53.2276 (.0826) | 0.9466 (.0007) | 0.8904 (.0007) | 1.5024 (.0008) | -41.5654 (.0650) | 0.7678 (.0006) | 0.8182 (.0006) | 1.6025 (.0009) |
| LSS | -49.6994 (.2140) | 1.0707 (.0027) | 0.7012 (.0034) | 1.4887 (.0009) | -41.1231 (.1593) | 0.6930 (.0013) | 0.8920 (.0020) | 1.5817 (.0009) |

Note: The asterisk * indicates the estimates are obtained from the bootstrap with a sample size of B=5,000. The values in parentheses are the corresponding standard errors of the estimates. For the model on the sons' heights, the estimates from all approaches illuminate that the contribution from the father to the height of his son is greater than that from the mother, because a larger regression coefficient means a larger change in the predicted variable given the same unit change in the regressor. In contrast, for the model on the daughters' heights, all the estimates except that from the LS clarify that the height of the daughter is mainly influenced by her mother rather than father. In the respect of the sum of regression efficiencies (SRE), the OR estimator, in general, fits Galton's data best, and the GMR as well as the LSS estimations are the near optimal ones, while the LS has the much lower SRE compared to the others.

*Table 3.  The* ASL *of hypotheses testing on the unequal regression coefficients*

| Regressions | Permutation Test | | Generalized bootstrap | |
|---|---|---|---|---|
| | $ASL_{Son}$ | $ASL_{Daughter}$ | $ASL_{Son}$ | $ASL_{Daughter}$ |
| OR | 0.000000 | 0.071953 | 0.000002 | 0.072099 |
| GMR | 0.063056 | 0.070482 | 0.062590 | 0.070239 |
| LSS | 0.000000 | 0.000022 | 0.000000 | 0.000079 |

Note: All ASLs are generated under the resampling size of 1,000,000. No apparent discrimination has been found between the ASLs from two different resampling procedures. As can be seen from the generalized bootstrap results, for the hypothesis test of the model on the sons' heights, the ASLs from both the OR and LSS approaches appear to be highly significant at significance level of $10^{-5}$, while that from the GMR is not that significant with ASL slightly smaller than 0.1. Meanwhile, referring to the model on the daughters' heights, the ASL from the LSS is again greatly significant at significance level of $10^{-4}$, while those from the OR and GMR approaches are weakly significant at significance level of 0.1.

*Table 4(a)  Simulation results of regression with two predictors for the large sample (n = 500) with small errors*

| | Small error, equal variances $\sigma_{11} = 4, \sigma_{22} = 4, \sigma_{\varepsilon\varepsilon} = 4$ | | | | Small error, unequal variances $\sigma_{11} = 4, \sigma_{22} = 8, \sigma_{\varepsilon\varepsilon} = 2$ | | | | Small error, unequal variances $\sigma_{11} = 2, \sigma_{22} = 4, \sigma_{\varepsilon\varepsilon} = 8$ | | | |
|---|---|---|---|---|---|---|---|---|---|---|---|---|
| | LS | OR | GMR | LSS | LS | OR | GMR | LSS | LS | OR | GMR | LSS |
| $\beta_0$ | 1.000092<br>0.023098<br>0.023098<br>Non-sig. | 0.999651<br>0.023765<br>0.023765<br>Non-sig. | 0.999653<br>0.023753<br>0.023753<br>Non-sig. | 0.999720<br>0.024335<br>0.024335<br>Non-sig. | 1.003270<br>0.027182<br>0.027171<br>$p < 0.05$ | 1.003699<br>0.028080<br>0.028066<br>$p < 0.05$ | 1.003699<br>0.028078<br>0.028065<br>$p < 0.05$ | 1.003867<br>0.028813<br>0.028798<br>$p < 0.05$ | 1.000410<br>0.027952<br>0.027952<br>Non-sig. | 1.000413<br>0.028937<br>0.028937<br>Non-sig. | 1.000410<br>0.028900<br>0.028900<br>Non-sig. | 1.000623<br>0.029673<br>0.029673<br>Non-sig. |
| $\beta_1$ | 0.961705<br>0.001693<br>0.000227<br>$p < 10^{-6}$ | 1.000227<br>0.000248<br>0.000248<br>Non-sig. | 1.000213<br>0.000224<br>0.000224<br>Non-sig. | 1.010509<br>0.000835<br>0.000725<br>$p < 10^{-6}$ | 0.961732<br>0.001727<br>0.000263<br>$p < 10^{-6}$ | 1.006579<br>0.000332<br>0.000289<br>$p < 10^{-6}$ | 1.005321<br>0.000285<br>0.000257<br>$p < 10^{-6}$ | 1.017386<br>0.001100<br>0.000797<br>$p < 10^{-6}$ | 0.980609<br>0.000655<br>0.000279<br>$p < 10^{-6}$ | 1.027427<br>0.001063<br>0.000310<br>$p < 10^{-6}$ | 1.026017<br>0.000952<br>0.000275<br>$p < 10^{-6}$ | 1.038880<br>0.002365<br>0.000853<br>$p < 10^{-6}$ |
| $\beta_2$ | 0.961355<br>0.001713<br>0.000220<br>$p < 10^{-6}$ | 0.999908<br>0.000240<br>0.000240<br>Non-sig. | 0.999916<br>0.000216<br>0.000216<br>Non-sig. | 1.010815<br>0.000851<br>0.000734<br>$p < 10^{-6}$ | 0.925834<br>0.005748<br>0.000248<br>$p < 10^{-6}$ | 0.967376<br>0.001339<br>0.000275<br>$p < 10^{-6}$ | 0.969399<br>0.001181<br>0.000244<br>$p < 10^{-6}$ | 0.979710<br>0.001202<br>0.000790<br>$p < 10^{-6}$ | 0.961811<br>0.001725<br>0.000266<br>$p < 10^{-6}$ | 1.006879<br>0.000345<br>0.000298<br>$p < 10^{-6}$ | 1.007244<br>0.000316<br>0.000264<br>$p < 10^{-6}$ | 1.019424<br>0.001209<br>0.000832<br>$p < 10^{-6}$ |
| TMSE | 0.026505<br>0.003406 | 0.024253<br>0.000488 | 0.024193<br>0.000440 | 0.026022<br>0.001686 | 0.034657<br>0.007475 | 0.029751<br>0.001671 | 0.029544<br>0.001466 | 0.031114<br>0.002302 | 0.030331<br>0.002379 | 0.030345<br>0.001408 | 0.030168<br>0.001268 | 0.033247<br>0.003574 |

Note: In the table, $\sigma_{11}$ is the variance of the measurement error in $X_1$, $\sigma_{22}$ is the variance of the measurement error in $X_2$, and $\sigma_{\varepsilon\varepsilon}$ is the variance of the measurement error in $Y$. For each regression coefficient estimate, the corresponding entries in each cell are the mean estimated value, the mean squared error, the variance, and the p-value of significance over the 10000 runs. For the total mean squared error (TMSE), the entries in each cell are TMSE of all coefficient estimates and the TMSE of only the two slope estimates.

Table 4(*b*)    Simulation results of regression with two predictors for the large sample (*n* = 500) *with large errors*

| | Large error, equal variances $\sigma_{11} = 36, \sigma_{22} = 36, \sigma_{\varepsilon\varepsilon} = 36$ | | | | Large error, unequal variances $\sigma_{11} = 36, \sigma_{22} = 49, \sigma_{\varepsilon\varepsilon} = 25$ | | | | Large error, unequal variances $\sigma_{11} = 25, \sigma_{22} = 36, \sigma_{\varepsilon\varepsilon} = 49$ | | | |
|---|---|---|---|---|---|---|---|---|---|---|---|---|
| | LS | OR | GMR | LSS | LS | OR | GMR | LSS | LS | OR | GMR | LSS |
| $\beta_0$ | 0.997827 0.182291 0.182286 Non-sig. | 1.000441 0.221954 0.221954 Non-sig. | 1.000471 0.221013 0.221013 Non-sig. | 1.000601 0.234266 0.234266 Non-sig. | 1.008014 0.166626 0.166562 $p < 0.05$ | 1.010110 0.201244 0.201141 $p < 0.05$ | 1.010286 0.203350 0.203244 $p < 0.05$ | 1.010793 0.214913 0.214796 $p < 0.05$ | 0.998550 0.195562 0.195560 Non-sig. | 1.000167 0.244593 0.244593 Non-sig. | 1.000042 0.239396 0.239396 Non-sig. | 1.000343 0.255804 0.255804 Non-sig. |
| $\beta_1$ | 0.734876 0.071605 0.001315 $p < 10^{-6}$ | 1.000154 0.002666 0.002666 Non-sig. | 1.000130 0.001268 0.001268 Non-sig. | 1.039081 0.004849 0.003321 $p < 10^{-6}$ | 0.734679 0.071644 0.001248 $p < 10^{-6}$ | 1.001561 0.002544 0.002541 $p < 0.005$ | 0.988737 0.001361 0.001235 $p < 10^{-6}$ | 1.030203 0.004189 0.003276 $p < 10^{-6}$ | 0.800491 0.041322 0.001518 $p < 10^{-6}$ | 1.127115 0.019474 0.003315 $p < 10^{-6}$ | 1.086537 0.008980 0.001491 $p < 10^{-6}$ | 1.141638 0.024042 0.003980 $p < 10^{-6}$ |
| $\beta_2$ | 0.735524 0.071260 0.001313 $p < 10^{-6}$ | 1.001219 0.002754 0.002753 $p < 0.05$ | 1.000749 0.001311 0.001310 $p < 0.05$ | 1.040393 0.005060 0.003429 $p < 10^{-6}$ | 0.671633 0.108971 0.001146 $p < 10^{-6}$ | 0.887540 0.014860 0.002213 $p < 10^{-6}$ | 0.920733 0.007383 0.001100 $p < 10^{-6}$ | 0.945169 0.006108 0.003101 $p < 10^{-6}$ | 0.736001 0.071131 0.001435 $p < 10^{-6}$ | 1.003175 0.003078 0.003067 $p < 10^{-6}$ | 1.016957 0.001692 0.001405 $p < 10^{-6}$ | 1.054379 0.006630 0.003672 $p < 10^{-6}$ |
| TMSE | 0.325156 0.142865 | 0.227374 0.005420 | 0.223592 0.002579 | 0.244174 0.009909 | 0.347241 0.180615 | 0.218648 0.017404 | 0.212095 0.008744 | 0.225209 0.010296 | 0.308016 0.112453 | 0.267144 0.022551 | 0.250068 0.010672 | 0.286475 0.030671 |

Note: In the table, $\sigma_{11}$ is the variance of the measurement error in $X_1$, $\sigma_{22}$ is the variance of the measurement error in $X_2$, and $\sigma_{\varepsilon\varepsilon}$ is the variance of the measurement error in *Y*. For each regression coefficient estimate, the corresponding entries in each cell are the mean estimated value, the mean squared error, the variance, and the p-value of significance over the 10000 runs. For the total mean squared error (TMSE), the entries in each cell are TMSE of all coefficient estimates and the TMSE of only the two slope estimates.